\documentclass{blois}

\def\be{\begin{equation}}
\def\ee{\end{equation}}
\def\bea{\begin{eqnarray}}
\def\eea{\end{eqnarray}}

\usepackage{amsmath,amssymb,slashed}

\def\be{\begin{equation}}
\def\ee{\end{equation}}
\def\bea{\begin{eqnarray}}
\def\eea{\end{eqnarray}}

\newcommand{\Photo}{}

\begin{document}
\title{Implications of heavy flavour measurements}

\author{Ulrich Haisch}

\address{Rudolf Peierls Centre for Theoretical Physics, University of Oxford \\ OX1 3PN Oxford, United Kingdom}

\maketitle\abstracts{A personal account of possible implications of recent heavy flavour measurements  is given.}

\section{Setting the stage}

The discovery of the Higgs boson by ATLAS\,\cite{Aad:2012tfa} and CMS\,\cite{Chatrchyan:2012xdj}  represents without doubt the highlight of  LHC Run I. All we have learnt in the last three years about the couplings of this new boson can be summarised in a single number,~i.e.~the combined best-fit signal strength\,\cite{Aad:2015gba,CMS:2014ega}
\begin{equation} \label{eq:1}
\mu_h= 1.1 \pm 0.1 \,.
\end{equation}
The $10\%$ agreement of $\mu_h$ with the Standard Model (SM) value $1$ allows to  constrain indirectly any beyond the SM (BSM) scenario.  For instance in the case of the coupling of the Higgs to two photons, one arrives at the following naive estimate of the corresponding signal strength 
\begin{equation}  \label{eq:2}
\raisebox{-1.1cm}{\includegraphics[width=4cm]{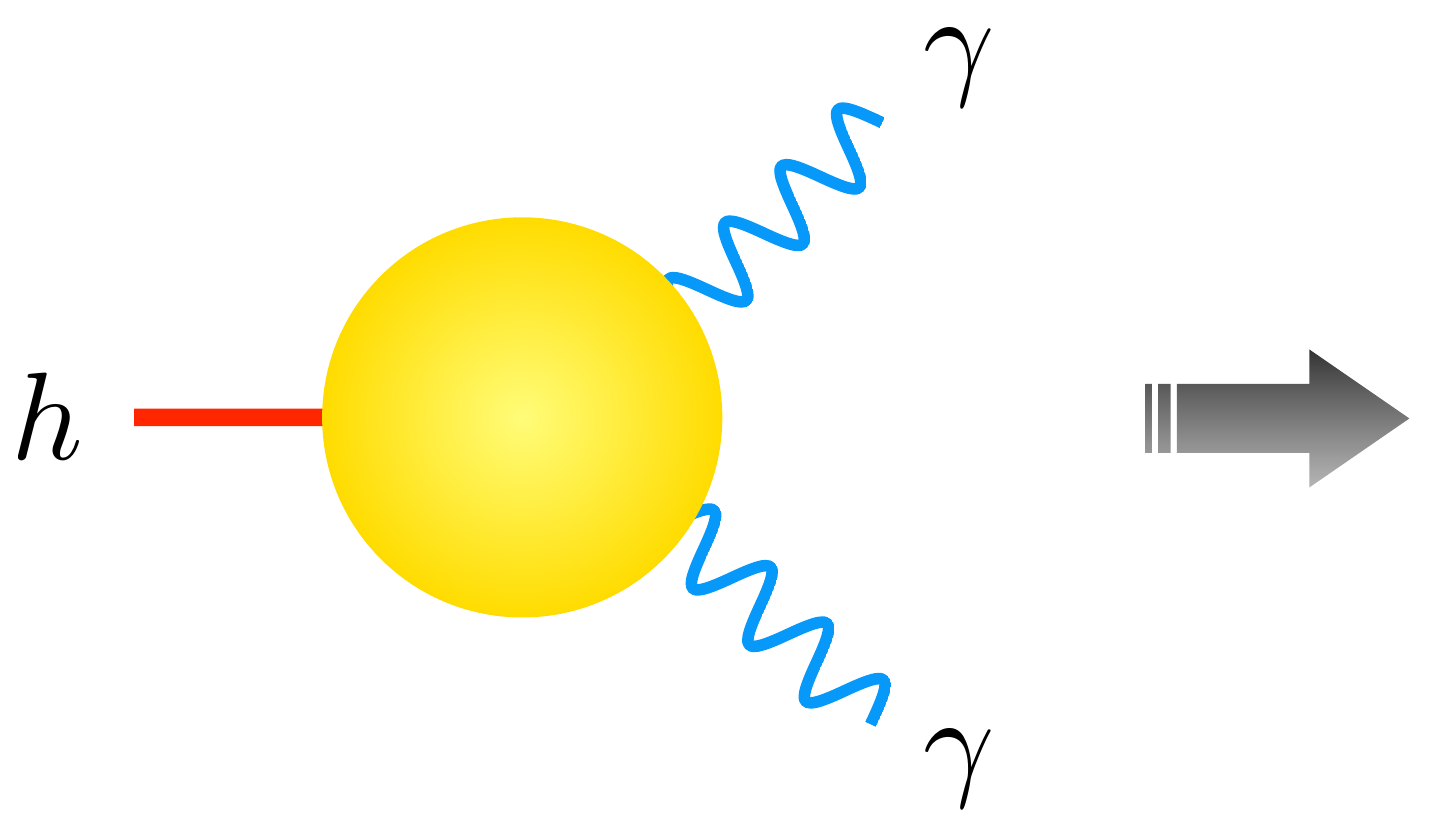} } \; \; \mu_{h \to \gamma \gamma} \simeq  1 \pm N  \frac{v^2}{\Lambda^2} \,.
\end{equation}
Here $v \simeq 246 \, {\rm GeV}$ is the Higgs vacuum expectation value, $\Lambda$ denotes the scale of new physics and $N$ parameterises our ignorance about the precise form of the BSM dynamics. Combining~(\ref{eq:1}) and~(\ref{eq:2}), one finds
\begin{equation} \label{eq:3}
\Lambda \gtrsim \sqrt{\frac{N}{10\%}} \, v \simeq \begin{cases}  0.8 \, {\rm TeV} \,, & N = 1 \,, \\[2mm]
3 \, {\rm TeV} \,, & N = 4 \pi  \,, \end{cases}
\end{equation}
where the first (second) case corresponds to a generic weakly-coupled (strongly-coupled) theory. In the best-case scenario, the LHC Run I measurements of the Higgs couplings  hence  allow to probe  new dynamics  in the few TeV regime. 

The flavour measurement in  Run I of the LHC that comes probably closest to the significance of the Higgs discovery is the observation of  the rare $B_s \to \mu^+ \mu^-$ decay.\,\cite{CMS:2014xfa} This  measurement leads to a signal strength\,\cite{Bobeth:2015zqa}  
\begin{equation} \label{eq:4}
\mu_{B_s \to \mu^+ \mu^-} =  0.78 \pm 0.18 \,, 
\end{equation}
which has a relative uncertainty of around $20 \%$. In order to translate (\ref{eq:4}) into a bound on $\Lambda$, we consider two specific BSM scenarios. The first case is that of a weakly-coupled $Z^\prime$ boson with generic flavour-changing tree-level quark couplings, while in our second benchmark we look at one-loop modifications of the $Z$ penguin assuming minimal-flavour violation (MFV).\,\cite{D'Ambrosio:2002ex} One estimates for the signal strengths in these two cases 
\begin{equation} \label{eq:5}
\begin{split}
& \raisebox{-0.95cm}{\includegraphics[width=3.35cm]{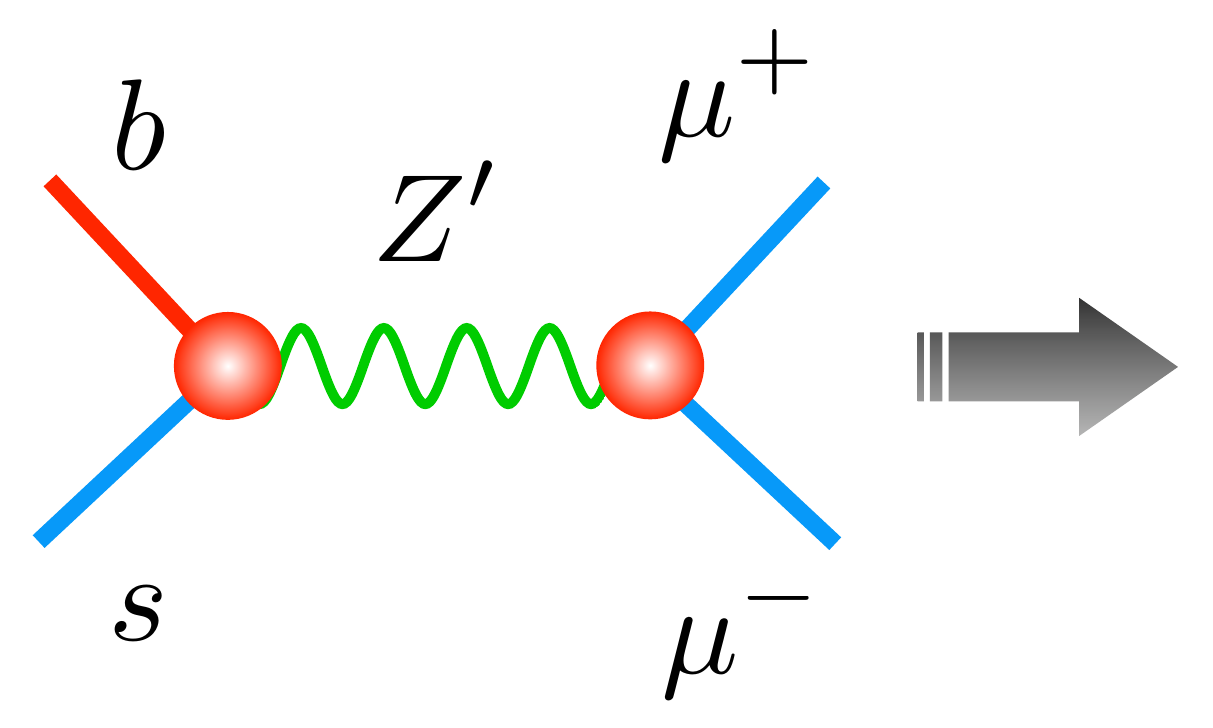} } \; \; \mu_{B_s \to \mu^+ \mu^-} \simeq  1 \pm \frac{4 \pi}{g^2 |V_{tb}^\ast V_{ts}|^2}  \frac{v^2}{\Lambda^2} \,, \\[1mm]
& \hspace{+5mm} \raisebox{-0.75cm}{\includegraphics[width=4.25cm]{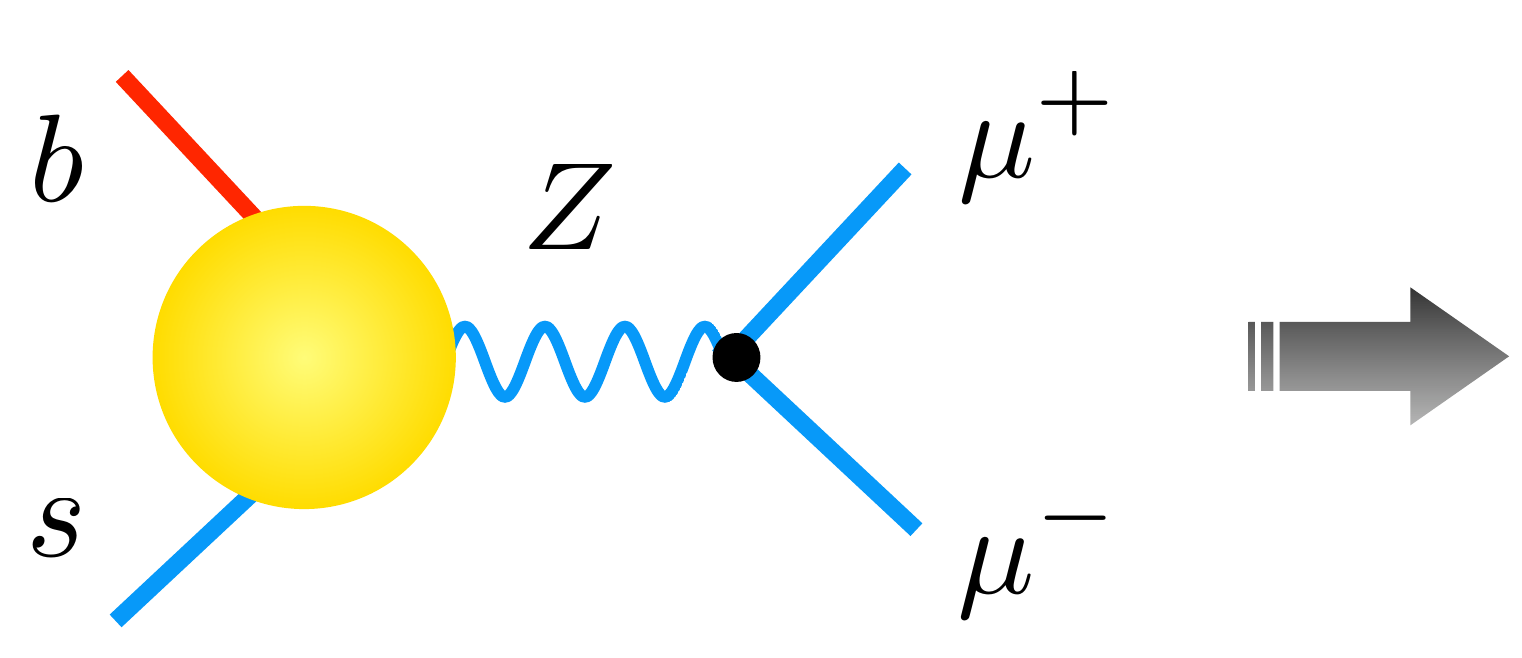} } \; \; \mu_{B_s \to \mu^+ \mu^-} \simeq  1 \pm  \frac{v^2}{\Lambda^2} \,,
\end{split}
\end{equation}
where $g \simeq 0.65$ is the $SU(2)_L$ coupling and $V_{ij}$ are the elements of the Cabibbo-Kobayashi-Maskawa matrix. From (\ref{eq:4}) and~(\ref{eq:5}) it then follows that 
\begin{equation}  \label{eq:6}
\Lambda \gtrsim \frac{v}{\sqrt{20\%}} \times \begin{cases} \displaystyle \frac{\sqrt{4 \pi}}{g |V_{tb}^\ast V_{ts}|} \\[4mm] 1 \end{cases} \!\!\! \!\!\simeq \begin{cases} 50 \, {\rm TeV} \,, \\[2mm] 0.6 \, {\rm TeV} \,.\end{cases} 
\end{equation}
The upshot of the above exercise is that even in the most pessimistic scenario,~i.e.~MFV, the LHC Run I sensitivity of  flavour observables to the new-physics scale $\Lambda$ is comparable to that of the Higgs couplings measurements by ATLAS and CMS. Like in the case of Higgs physics, we are now in era of precision physics for what concerns quark flavour. Further progress is therefore likely to depend on how well experimentalists can measure and how well theorists can predict --- of course, there is still room for surprises!

\section{Flavour precision tests}

A simple but educated example which shows that flavour physics has indeed entered a new  era is provided by the comparison of the constraints on certain BSM contributions  from $B_s \to \mu^+ \mu^-$ with that  determined from electroweak precision observables (EWPOs). In fact,  in a wide class of models such as MFV or partial compositeness the most important deviations from the SM in $B_s \to \mu^+ \mu^-$ and $Z \to b \bar b$ can be described in terms of modified $Z$-boson couplings at zero-momentum transfer\,\cite{Haisch:2007ia}
\begin{equation} \label{eq:7}
{\cal L} \supset \frac{e}{s_w c_w}  \, V_{ti}^\ast V_{tj} \, \delta g_L \, \bar d_L^i \slashed{Z}  d_L^j \,, 
\end{equation}
with $s_w$ ($c_w$) the sine (cosine) of the weak mixing angle and $\delta g_L$ a flavour-blind coefficient. This universal coefficient enters the signal strength for $B_s \to \mu^+ \mu^-$ in the following way\,\cite{Bobeth:2015zqa}  
\begin{equation} \label{eq:8}
\mu_{B_s \to \mu^+ \mu^-} \simeq ( 1 + 200 \, \delta g_L)^2 \,,
\end{equation}
and also shifts the left-handed $Z b \bar b$ coupling from its SM value $g_L^b \simeq -1/2 + s_w^2/3$. Utilising (\ref{eq:4}) as well as the results of a recent global analysis of  EWPOs,\,\cite{Ciuchini:2013pca} one obtains 
\begin{equation} \label{eq:9}
\delta g_L = \begin{cases} (-0.6 \pm 0.5) \cdot 10^{-3} \,, &  {\rm from \;} B_s \to \mu^+ \mu^- \,, \\[2mm]
 (1.6 \pm 1.5) \cdot 10^{-3} \,, & {\rm from \;} Z \to b \bar b \,. \end{cases}
 \end{equation}
These numbers show clearly that the experimental precision reached on the branching ratio of~$B_s \to \mu^+ \mu^-$ is such that this observable sets the dominant constraints on possible modified~$Z$-boson couplings. In this sense, $B_s \to \mu^+ \mu^-$ can now be regarded as a EWPO.\,\cite{Guadagnoli:2013mru} Pre LHC, this was not the case since the constraints from $Z \to b \bar b$ were stronger than those arising from all the~$b \to s Z$ and  $s \to d Z$ transitions.\,\cite{Haisch:2007ia,Bobeth:2005ck}

\begin{figure}[!t]
\begin{center}
\raisebox{1.75cm}{\includegraphics[width=0.5 \textwidth]{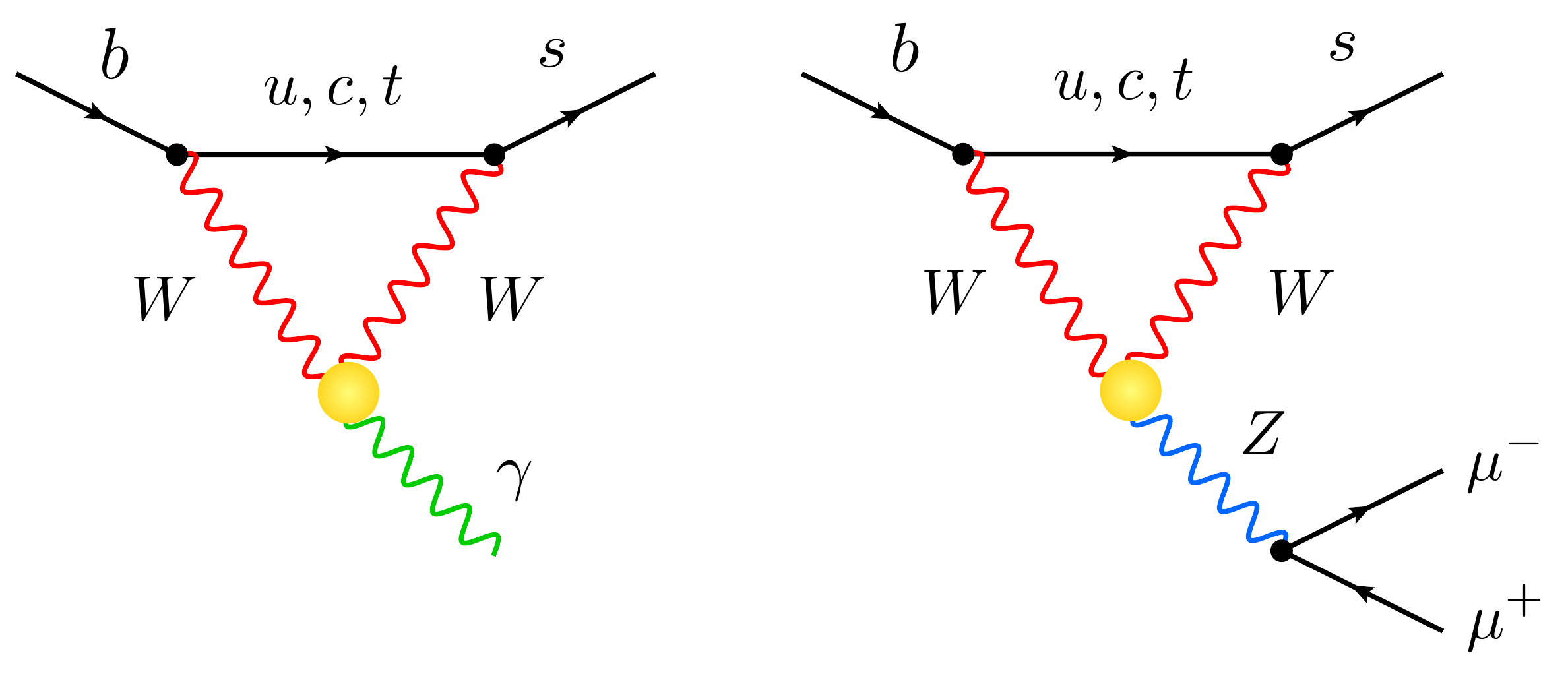} } \qquad 
\includegraphics[width=0.425 \textwidth]{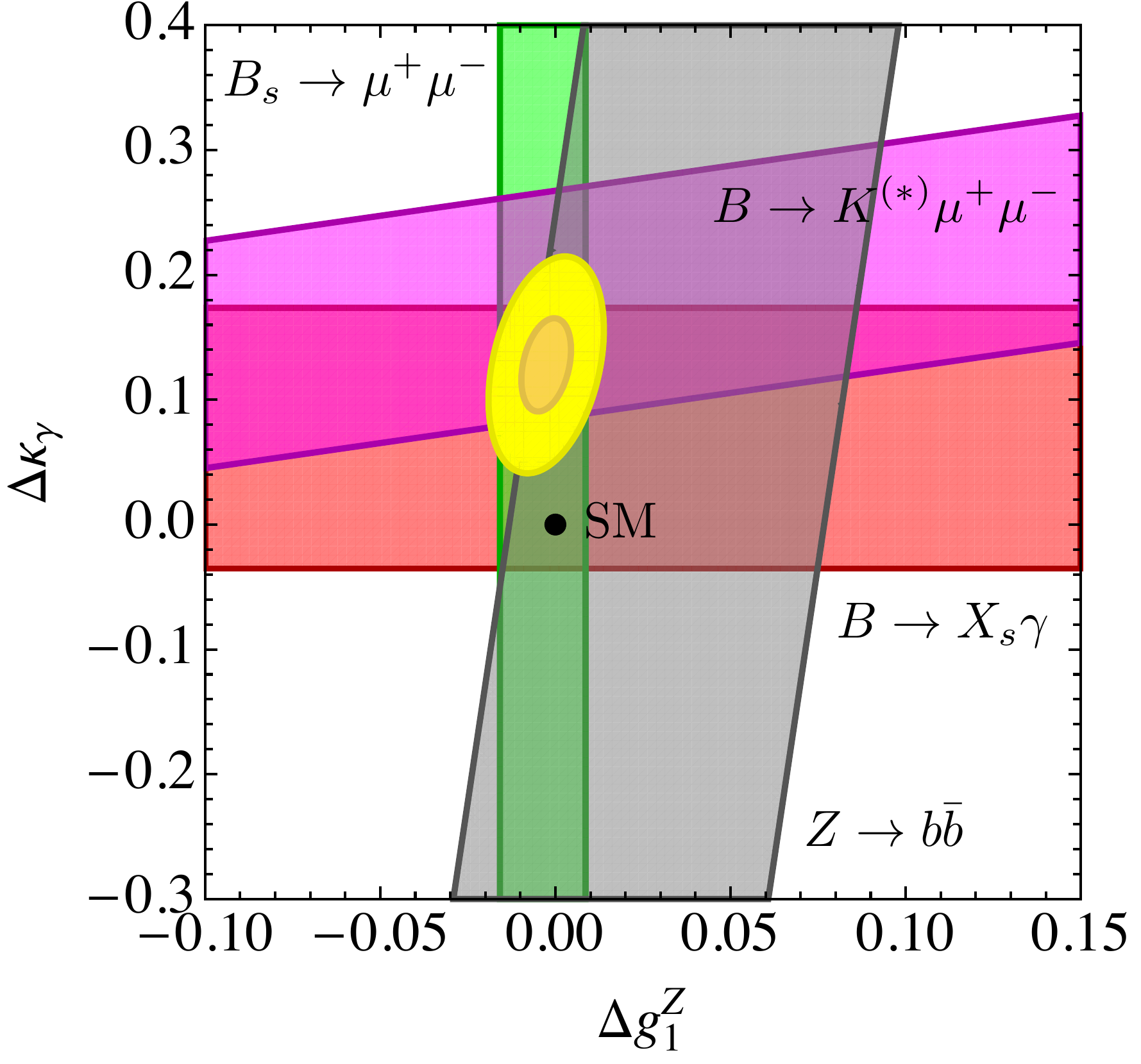} 
\end{center}
\vspace{-4mm}
\caption{\label{fig:1} Left: Examples of one-loop diagrams that generate a $b \to s \gamma$ and a $b \to s \mu^+ \mu^-$ transition.  The TGCs are indicated by yellow circles, while the SM vertices are represented by black dots. Right: Allowed regions in the $\Delta g_1^Z \hspace{0.25mm}$--$ \hspace{0.25mm} \Delta \kappa_\gamma$ plane. The red, magenta, green and grey contours correspond to the~68\% confidence level (CL) best fit regions following from  $B \to X_s \gamma$, $B \to K^{(\ast)} \mu^+ \mu^-$, $B_s \to \mu^+ \mu^-$ and $Z \to b \bar b$. The yellow and orange contours show the 68\% CL and 95\% CL regions arising from a combination of the individual constraints, while the black point correspond to the SM.}
\end{figure}

Under motivated assumptions about the underlying flavour structure, quark-flavour observables are also sensitive probes of triple gauge boson couplings (TGCs). These interactions are commonly parameterised as\,\cite{Hagiwara:1986vm} 
\begin{equation} \label{eq:10}
\begin{split}
{\cal L}_{WWV} = -i g_{WWV} \, \bigg  [ & \hspace{0.25mm} \left (1 + \Delta g_1^V \right ) \left ( W_{\mu \nu}^+  \hspace{0.25mm} W^{-  \hspace{0.25mm} \mu}  \hspace{0.25mm} V^\nu - W_\mu^+  \hspace{0.25mm} V_\nu  \hspace{0.25mm} W^{-  \hspace{0.25mm} \mu \nu}  \right ) \\[2mm] & + \left (1 + \Delta \kappa_V \right )  \hspace{0.25mm}  W_\mu^+  \hspace{0.25mm}  W_\nu^-  \hspace{0.25mm}  V^{\mu \nu} + \frac{\lambda_V}{m_W^2} \, W_{\mu \nu}^+ \hspace{0.25mm} W^{- \hspace{0.25mm} \nu \rho} \hspace{0.25mm} {V_\rho}^\mu \hspace{0.25mm} \bigg ] \,,
\end{split}
\end{equation}
with $V=\gamma, Z$. The overall coupling strengths are defined by $g_{WW\gamma} = g s_w = g^\prime c_w= e$ and $g_{WWZ} = g c_w$, where $W_{\mu \nu}^{\pm} = \partial_\mu W_\nu^\pm - \partial_\nu W_\mu^\pm$ and $V_{\mu \nu} =   \partial_\mu V_\nu - \partial_\nu V_\mu$ with $W_\mu^\pm$ and $V_\mu$ referring to the physical gauge boson fields. Furthermore, $\Delta g_1^\gamma = 0$ as a result of gauge invariance and
\begin{equation} \label{eq:11}
 \Delta \kappa_Z = \Delta g_1^Z - \frac{s_w^2}{c_w^2} \, \Delta \kappa_\gamma \,, \qquad \lambda_Z = \lambda_\gamma \,,
\end{equation}
if only dimension-6 contributions are considered.\,\cite{Hagiwara:1993ck} The Lagrangian introduced in  (\ref{eq:10}) induces contributions to radiative and rare $B$ decays, kaon physics and as well as the decay $Z \to b \bar b$, meaning that the coefficients $\Delta g_1^Z$, $\Delta \kappa_\gamma$ and $\lambda_\gamma$ can be constrained from this data. The Feynman graphs that give rise to the modifications in $b \to s \gamma$ and $b \to s \mu^+ \mu^-$ are depicted on the left-hand side in Figure~\ref{fig:1}. 

Employing the results of the recent analysis  of radiative  and rare $b \to s$ observables\,\cite{Altmannshofer:2014rta} together with the information arising from a global fit to the EWPOs,\,\cite{Ciuchini:2013pca} one obtains the constraints on the $\Delta g_1^Z \hspace{0.25mm}$--$ \hspace{0.25mm} \Delta \kappa_\gamma$ plane as displayed on the right in Figure~\ref{fig:1}. For $\lambda_\gamma = 0$, the allowed parameter ranges at the~68\% CL read\,\cite{Bobeth:2015zqa}   
\begin{equation} \label{eq:12}
\Delta g_1^Z = -0.003 \pm 0.007 \,, \qquad \Delta \kappa_\gamma = 0.13 \pm 0.04 \,.
\end{equation}
These fit results  should be contrasted  with the limits that can be derived from electroweak gauge boson pair production at LEP~II, the Tevatron and the LHC as well as Higgs physics. For instance,  the very recent global analysis of TGCs\,\cite{Falkowski:2015jaa} obtains the following~68\%~CL bounds 
\begin{equation} \label{eq:13}
\Delta g_1^Z = 0.017 \pm 0.023 \,,  \qquad \Delta \kappa_\gamma = 0.047 \pm 0.034 \,, \qquad \lambda_\gamma = -0.089 \pm 0.042 \,.
\end{equation}
We see that compared to (\ref{eq:12}) the global constraint on $\Delta g_1^Z$ from LEP~II and Higgs data are notable weaker, while in the case of the parameter $\Delta \kappa_\gamma$ the uncertainties in (\ref{eq:12}) and (\ref{eq:13}) are similar. These findings illustrate that precision measurements of $B \to K^{(\ast)} \mu^+ \mu^-$ and $B_s \to \phi \mu^+ \mu^-$, possible at LHCb, provide another powerful probe of electroweak physics. Notice also that in (\ref{eq:12}) the best fit  point for $\Delta \kappa_\gamma$  is by about $3 \sigma$ away from the SM as a result of the various deviations seen in rare $b \to s \ell^+ \ell^-$ transitions. 

Before discussing these hints of BSM physics in more detail, let me add that also in the case of anomalous $Z t \bar t$ couplings the present indirect bounds from flavour and electroweak precision physics\,\cite{Brod:2014hsa} are more stringent than the direct bounds that the LHC might be able to set at high luminosities.\,\cite{Rontsch:2014cca} Still direct tests of both the  TGCs and anomalous $Z t \bar t$ couplings  have to be undertaken at the LHC, since in contrast to the indirect test they probe the relevant interactions at tree level. On the other hand, one should also not forget that indirect probes do exist and can add valuable and complementary informations to the high-$p_T$ measurements. 

\section{Anomalies in the flavour sector}

There are several anomalies in quark-flavour physics that exceed the level of $2\sigma$. The list includes $B \to K^{(\ast)} \mu^+ \mu^-$, $B_s \to \phi \mu^+ \mu^-$, $R_K$, $B \to D^{(\ast)} \tau \nu$, $V_{ub}$, $V_{cb}$, the dimuon CP asymmetry and $\epsilon^\prime/\epsilon$. Given the space limitations I am facing, I will  in the following only discuss the anomalies seen by LHCb in the $b \to s \ell^+ \ell^-$ channels. 

The first hint of a possible sizeable BSM  contribution in $b \to s$ transitions dates back  more then two years\,\cite{Descotes-Genon:2013wba} when LHCb presented their first results on the angular distributions in $B \to K^\ast \mu^+ \mu^-$.\,\cite{Aaij:2013qta} These results were based on $1 \, {\rm fb}^{-1}$ of $7 \, {\rm TeV}$ data and showed a  deviation with a local significance of $3.7 \sigma$  in one of the angular observables called $P_5^\prime$\,\cite{DescotesGenon:2012zf} for dimuon invariant masses $q^2 \in [4.30,8.68] \, {\rm GeV}^2$. Using the full LHCb Run I data sample of $3 \, {\rm fb}^{-1}$, the angular analysis of the $B \to K^\ast \mu^+ \mu^-$ decay has been updated recently.\,\cite{LHCb:2015dla} The improved measurements are  in good agreement with the earlier  results and confirm the anomaly seen before in the $P_5^\prime$ distribution. Deviations of $2.9 \sigma$ are now found in  two bins with $q^2 \in [4,6] \, {\rm GeV}^2$ and $q^2 \in [6,8] \, {\rm GeV}^2$.  A second type of deviations concerns the LHCb branching ratio measurements of $B_s \to \phi \mu^+ \mu^-$\,\cite{Aaij:2015esa} and $B \to K^{(\ast)} \mu^+ \mu^-$,\,\cite{Aaij:2014pli} which are all low compared the latest  lattice QCD\,\cite{Horgan:2013pva} and light-cone sum rule~(LCSR) predictions.\,\cite{Straub:2015ica}  The most significant deviation is found  in the $B_s \to \phi \mu^+ \mu^-$  channel for $q^2 \in [1, 6] \, {\rm GeV}^2$ and amounts to $3.3 \sigma$. A third piece of the puzzle is provided by a possible sign of lepton-flavour non-universality in $B \to K \ell^+ \ell^-$,\,\cite{Aaij:2014ora}
\begin{equation} \label{eq:14}
R_K^{q^2 \in [1, 6] \, {\rm GeV}^2} = \frac{\int_{1 \, {\rm GeV}^2}^{6 \, {\rm GeV}^2} dq^2 \, \frac{d\Gamma (B \to K \mu^+ \mu^-  )}{dq^2}}{\int_{1 \, {\rm GeV}^2}^{6 \, {\rm GeV}^2} dq^2 \, \frac{d\Gamma (B \to K e^+ e^- )}{dq^2}} = 0.745^{+0.090}_{-0.075} \pm  0.036\,,
\end{equation}
which deviates by $2.6 \sigma$ from the SM prediction $R_K^{q^2 \in  [1, 6] \, {\rm GeV}^2} \simeq 1$. The observables $R_K$, $P_5^\prime$ and the differential rates in $B \to K^{(\ast)} \mu^+ \mu^-$, $B_s \to \phi \mu^+ \mu^-$ are plagued by quite different systematic errors of both experimental and theoretical origin. On the theory side one should worry about electromagnetic effects ($R_K$),\,\cite{LHCb14talk} form factor uncertainties, power corrections, long-distance $c \bar c$ effects and violation of quark-hadron duality ($P_5^\prime$,  $B \to K^{(\ast)} \mu^+ \mu^-$, $B_s \to \phi \mu^+ \mu^-$).  A better understanding of all these issues is certainly required to fully exploit the existing as well as the upcoming LHCb data. Instead of dwelling on these problems, I will discuss now what the data might tells us given our present theoretical understanding of radiative and rare $b \to s$ transitions. 

\begin{figure}[!t]
\begin{center}
\raisebox{0.5cm}{\includegraphics[width=0.475 \textwidth]{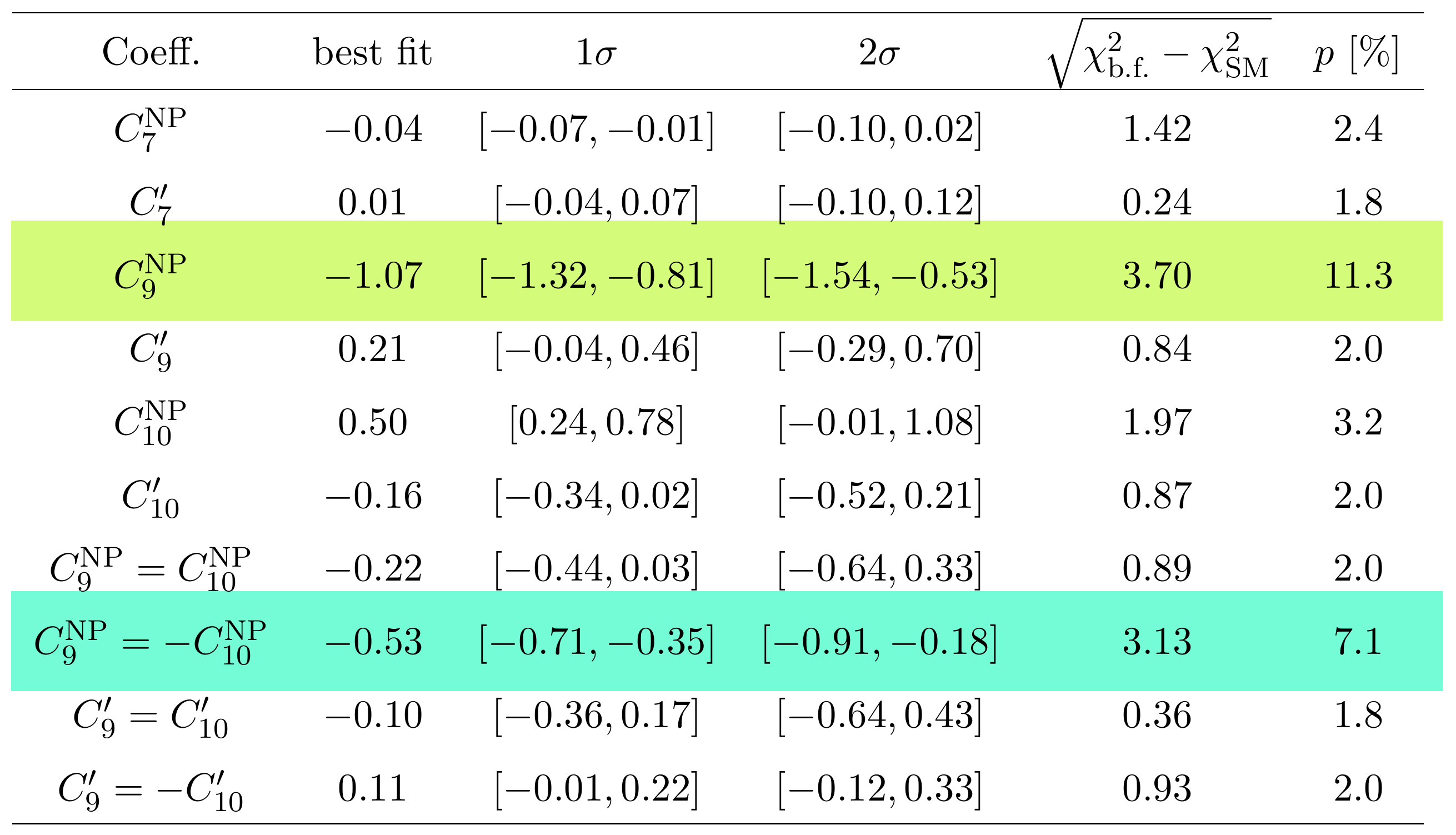} } \quad 
\includegraphics[width=0.475 \textwidth]{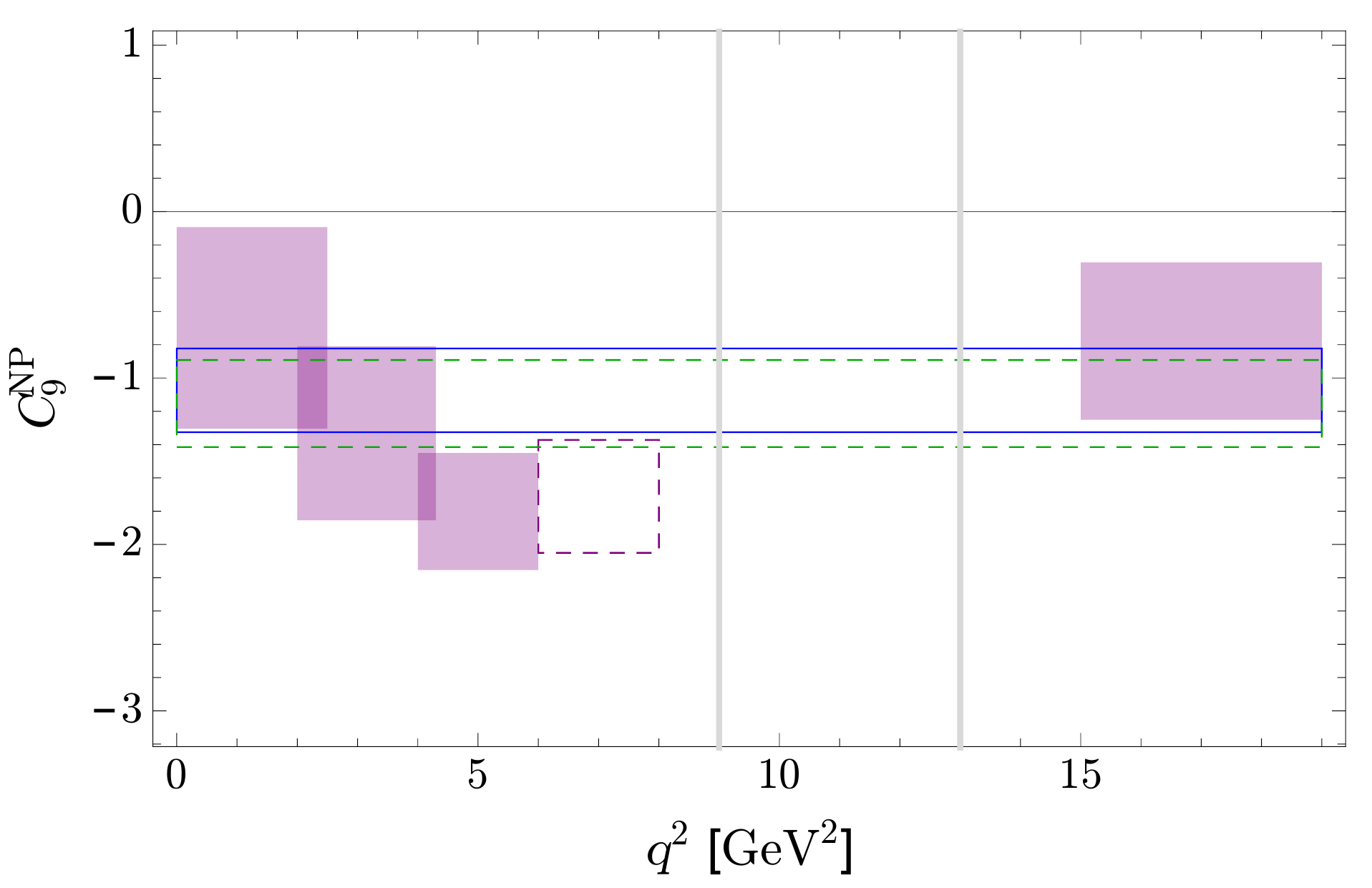} 
\end{center}
\vspace{-4mm}
\caption{\label{fig:2} Left:  Constraints on individual Wilson coefficients, assuming them to be real. The scenarios with the two highest $p$ values are highlighted in colour. Right: Ranges preferred at $1 \sigma$ for a new-physics contribution to $C_9$ from a bin-wise fit to all $B \to K^\ast \mu^+ \mu^-$ observables (purple). For comparison the $1 \sigma$ bands following  from the global fit (blue) and  from a fit to only $B \to K^\ast \mu^+ \mu^-$ observables  (green) are also shown. The vertical grey lines indicate the location of the $J/\psi$ and $J/\psi^\prime$ resonances.}
\end{figure}

Performing a global fit to 88 different $b \to s$ observables the recent work\,\cite{Altmannshofer:2015sma} derives model-independent constraints on  BSM scenarios that can be described in an effective field theory language. Assuming that all the Wilson coefficients are real (only the time-dependent CP asymmetry in $B \to K^\ast \gamma$ has been measured, leaving the imaginary parts of the Wilson coefficients essentially unconstrained) and considering one at a time, one obtains the results shown in the table on the left in Figure~\ref{fig:2}. The results with the two highest $p$ values are $C_9^{\rm NP} \sim -1$~($p = 11.3\%$) and $C_9^{\rm NP} = -C_{10}^{\rm NP} \sim -0.5$ ($p = 7.1\%$). The former solution  correspond to a relative  shift of ${\cal O} (-25\%)$ in the Wilson coefficient of the semi-leptonic vector operator, while in the latter case  both the Wilson coefficients of the semi-leptonic vector and axial-vector operator are modified simultaneously by ${\cal O}(-10\%)$. All other scenarios do not  improve notable upon the SM~($p =2.1\%$). 

The upshot of the global analysis\,\cite{Altmannshofer:2015sma} is thus that there are two simple scenarios of BSM physics that are preferred over the SM by more than $3 \sigma$ (which is a non-trivial feature), but it is also fair to say that no solution really nails it,~i.e.~leads to a very good description of all data. Moreover, the finding that the best fit corresponds to a modification of the Wilson coefficient of the semi-leptonic vector operator is a bit worrisome: long-distance $c \bar c$ effects mediated by virtual photon exchange also have a vector-like coupling to leptons and thus could mimic a BSM effect in the Wilson coefficient $C_9$. In fact, based on the existing data the possibility that some of the deviations are not due to BSM physics but a result of unaccounted  hadronic effects can already be tested.\,\cite{Altmannshofer:2015sma,Descotes-Genon:2015uva} With the finer binning of the latest $B \to K^\ast \mu^+ \mu^-$ analysis,\,\cite{LHCb:2015dla} one is now able to determine the preferred range of a hypothetical BSM contribution to say $C_9$ separately in each  $q^2$-bin. The outcome of such an exercise\,\cite{Altmannshofer:2015sma} is displayed on the right-hand side of Figure~\ref{fig:2}. The values of the BSM contribution to $C_9$ preferred by a bin-wise fit to all $B \to K^\ast \ell^+ \ell^-$ data is indicated in purple, while the blue (green) band corresponds to the $1 \sigma$ region following  from the global fit  (fit to only $B \to K^\ast \mu^+ \mu^-$ observables). A comparison of the different results in fact  allows to shed some light on the possible origin of the observed anomalies, because short-distance new physics should lead to a $q^2$-independent shift in $C_9$, whereas long-distance effects are expected to have a non-trivial $q^2$ dependence. While at the $1 \sigma$ level the purple band is indeed consistent with being a straight line, one cannot help but notice that the preferred fit values for $C_9^{\rm NP}$ grow in magnitude when approaching the $J/\psi$ resonance  from below in $q^2$. Qualitatively, this is the behaviour expected from a non-factorisable $c \bar c$ contribution, but making any quantitive statement is notoriously difficult given our limited understanding of soft QCD. Based on the existing model calculation using LCSRs,\,\cite{Khodjamirian:2010vf} the possibility that part of the deviations seen in $P_5^\prime$ in the $q^2 \in [4, 8] \, {\rm GeV}^2$ range is due to long-distance charm-loop effects interfering destructively with the SM  can certainly not be  excluded\,\cite{LHCb14talk,Descotes-Genon:2014uoa} --- even a full resolution is possible but relies on extrapolating the prediction of a Breit-Wigner resonance model to $q^2$ values far away from the resonance peaks.\,\cite{Lyon:2014hpa} In view of this unclear situation, any theoretical or experimental idea that would for instance allow to pin down the interference pattern of the short- and long-distance contributions below the $J/\psi$ resonances is very welcome. 

What are the possible new-physics implications of the $C_9^{\rm NP} \sim -1$ fit solution? Since the minimal supersymmetric SM, simple realisations of compositeness and minimal lepto-quark scenarios 
lead either to $|C_9^{\rm NP}| \ll |C_{10}^{\rm NP}|$ or $C_9^{\rm NP} =  \pm C_{10}^{\rm NP}$, these models fail to give the solution of the global fit with the highest $p$ value. The observed deviation can be addressed in $Z^\prime$-boson models that have vector-like couplings to muons. Two types of such scenarios have been discussed in the literature: the first class of theories are based on a $SU(3)_L \times U(1)_X$ symmetry,\,\cite{Gauld:2013qja,Buras:2013dea} while the second class of models are built around a $U(1)_{L_{\mu} - L_{\tau}}$ symmetry.\,\cite{Altmannshofer:2014cfa,Crivellin:2015mga,Crivellin:2015lwa} Hereafter these two classes of models will be called 3-3-1 and $L_{\mu} - L_{\tau}$, respectively. 

\begin{figure}[!t]
\begin{center}
\includegraphics[width=0.975 \textwidth]{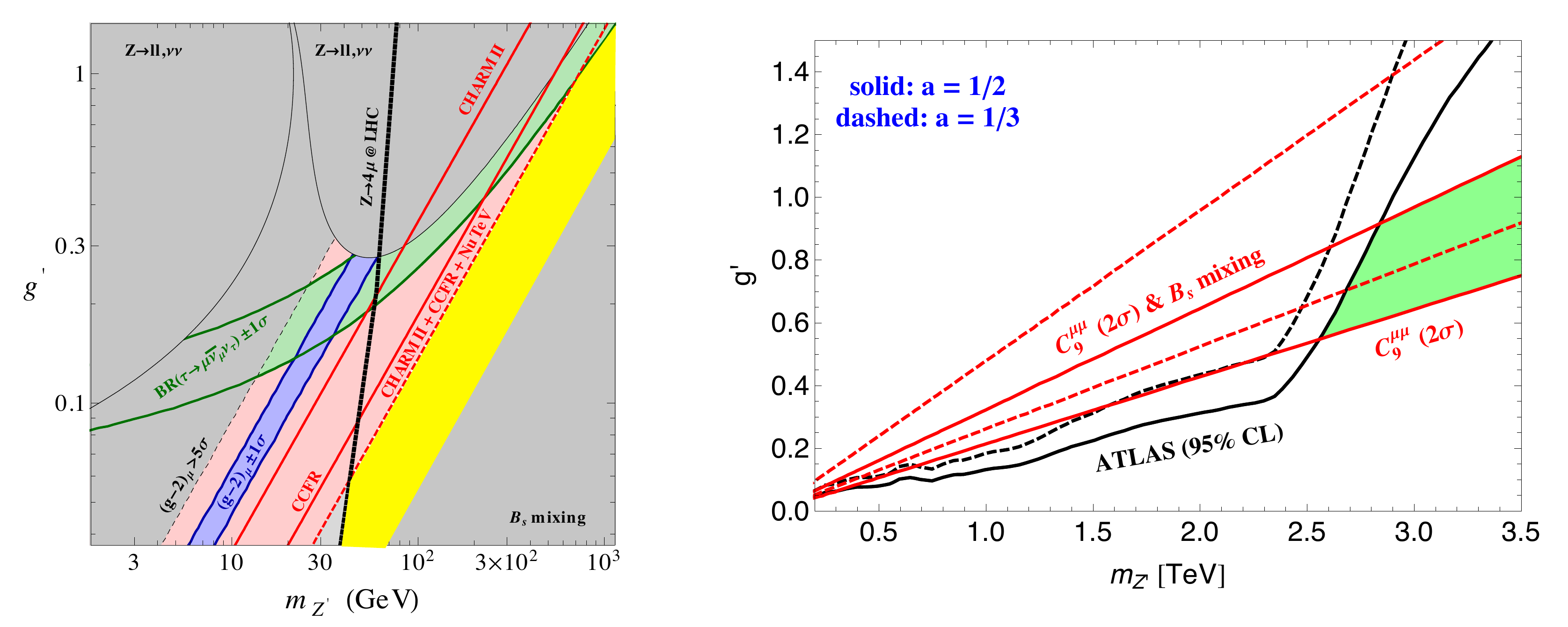} 
\end{center}
\vspace{-4mm}
\caption{\label{fig:3} Left:  Constraints on the model parameter space arising from the anomalous magnetic moment of the muon (blue),  $\tau \to \mu \nu_\tau \bar \nu_\mu$ (green), neutrino trident production (red), $pp \to Z \to 4 \mu$ (black), the decay widths of the $Z$ boson (grey) and $B_s$--$\bar B_s$ mixing (grey). In the yellow shaded region  all constraints are satisfied and the anomalies in the $b \to s \ell^+ \ell^-$ transitions are addressed.  Right: Constraints in the $M_{Z^\prime}\,$--$\,g^\prime$ plane arising from dimuon resonance searches at ATLAS for two different assignments of $U(1)^\prime$ charges (black curves). The green shaded region indicates the parameter space which is allowed by $pp \to Z^\prime \to \mu^+ \mu^-$ and $B_s$--$\bar B_s$ mixing and favoured by the $b \to s$ data. }
\end{figure}

In 3-3-1 models the $Z^\prime$-boson  coupling to leptons can be made almost vector-like by a suitable choice of charge normalisation and the $s \bar b Z^\prime$ coupling can be arranged to be MFV-like by 
alignment in the up-type quark sector.\,\cite{Gauld:2013qja} In order to obtain  $C_9^{\rm NP} \sim -1$ the mass of the $Z^\prime$ boson has to lie in the range of ${\cal O} (8 \, {\rm TeV})$. For such large values of $M_{Z^\prime}$ all other constraints following for instance from $B_s$--$\bar B_s$ mixing, atomic parity violation, unitarity of the quark mixing matrix as well as contact-interactions limits from LEP and direct $Z^\prime$ search bounds from LHC Run I are avoided.\,\cite{Gauld:2013qja,Buras:2013dea}  Since the $Z^\prime$ boson couples universally to the charged leptons in 3-3-1 models the $R_K$ 
anomaly (\ref{eq:14}) cannot be explained in these types of BSM scenarios. Furthermore, the minimal 3-3-1 model that can address the $P_5^\prime$ anomaly has a Landau-like pole in the $U(1)_X$ coupling at ${\cal O} ( 4 \, {\rm TeV})$ and thus needs to be extended to render a viable solution. One possible extension consists in adding leptonic triplets to the model, which has the further asset that in such a setup small neutrino masses can be generated in a natural way via an inverse seesaw mechanism.\,\cite{inpreparation}

$Z^\prime$-boson models in which the difference between the muon- and tau-lepton number $L_\mu - L_\tau$ is gauged provide a solution to both the $P_5^\prime$ and $R_K$ anomaly. This gauging automatically leads to a muonic (and tauonic) vector current, while not inducing a $Z^\prime$-boson coupling to electrons at tree level. The required left-handed $s \bar b Z^\prime$ coupling can either be obtained by mixing the SM quarks with vector-like matter\,\cite{Altmannshofer:2014cfa,Crivellin:2015mga} or by introducing appropriate horizontal gauge symmetries.\,\cite{Crivellin:2015lwa} In the first case the couplings to first generation quarks can be dialled to be small, which allows to avoid the stringent LHC bounds from $p p \to Z^\prime \to \mu^+ \mu^-$. Still the parameter space of such models is subject to a variety of constraints that however can all be fulfilled if $M_{Z^\prime} \gtrsim 40 \, {\rm GeV}$ and $M_{Z^\prime}/g^\prime \sim {\rm TeV}$.\,\cite{Altmannshofer:2014cfa} This is illustrated in the left panel of Figure~\ref{fig:3}. In contrast, in models with an additional  horizontal $U(1)^\prime$ symmetry the constraints from Drell-Yan $Z^\prime$-boson production can generically not be dodged.\,\cite{Crivellin:2015lwa}  The constraints in the $M_{Z^\prime}\, $--$\, g^\prime$ plane that arise from the ATLAS resonance search in the dimuon channel\,\cite{Aad:2014cka} are shown on the right-hand side in Figure~\ref{fig:3}. One observes that  in horizontal $L_\mu-L_\tau$ models employing the present $p p \to Z^\prime \to \mu^+ \mu^-$ bounds restricts the allowed parameter space to  $M_{Z^\prime} \gtrsim 2.5 \, {\rm TeV}$ and $g^\prime \gtrsim 0.5$, rather independent of the precise choice of $U(1)^\prime$ charges. 

\section{Electroweak physics}

\begin{figure}[!t]
\begin{center}
\includegraphics[width=0.95 \textwidth]{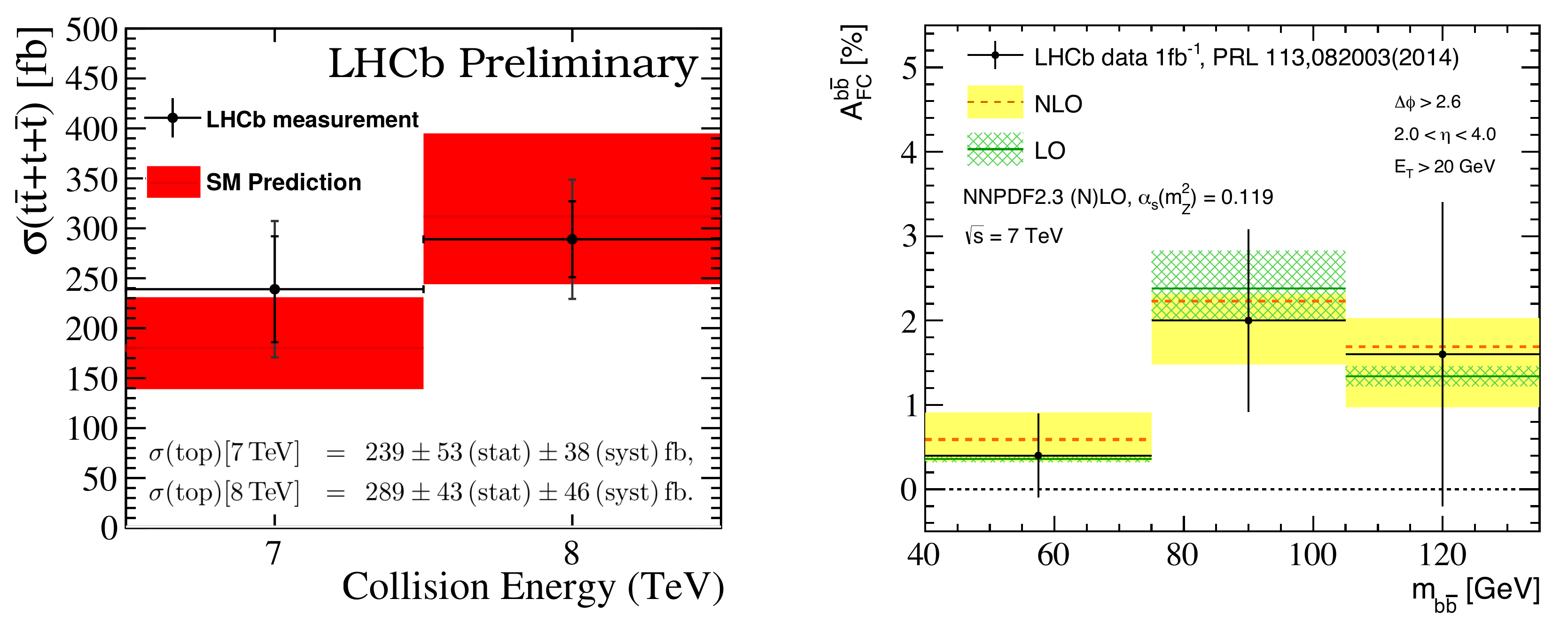} 
\end{center}
\vspace{-4mm}
\caption{\label{fig:4} Left:   Comparison of the LHCb measurements at $7 \, {\rm TeV}$ and $8 \, {\rm TeV}$ of the sum of the production cross sections of top-quark pairs and single-top (black error bars) to the corresponding SM expectations (red bands). Right: NLO  and leading order (LO) predictions of the beauty-quark forward-central asymmetry at $7 \, {\rm TeV}$ within the LHCb acceptance (yellow and green). The statistical and systematic uncertainties of the measurement have been added in quadrature to obtain the shown experimental error bars (black).}
\end{figure}

It is worth recalling that LHCb has a rich programme beyond pure quark-flavour physics as reflected by the activities in both the QCD, electroweak and exotica and the jets working group. One recent highlight of this programme is the first observation of top-quark production in the forward region.\,\cite{Coco,Aaij:2015mwa} As illustrated in the left panel in Figure~\ref{fig:4}, the cross-section results for the sum of top-quark pair and single-top production are in good agreement with the next-to-leading order (NLO) QCD predictions of  $(180^{+51}_{-41}) \, {\rm fb}$ $\big ($$(312^{+83}_{-68}) \, {\rm fb}$$\big)$ at $7 \, (8) \, {\rm TeV}$. The differential distributions of the yield and charge asymmetry are also consistent with the SM expectations. These measurements are not only of general importance as SM tests but of considerable theoretical interest. For instance, the enhancement at forward rapidities of $t \bar t$ production via $q \bar q$ and $q g$ scattering, relative to $gg$ fusion, can result in larger charge top-quark pair asymmetries. This feature gives LHCb unique abilities to probe BSM physics in the top-quark sector.\,\cite{Kagan:2011yx,Gauld:2014pxa} In addition, forward top-quark production can be used to constrain the gluon parton distribution function, which in turn results in reduced theoretical uncertainties for many SM processes.\,\cite{Gauld:2013aja}

While measurements of asymmetric $t \bar t$ production have not been performed at LHCb in Run~I, a first measurement of the angular asymmetry in  bottom-quark pair production based on $1 \, {\rm fb}^{-1}$ of $7 \, {\rm TeV}$ data was possible.\,\cite{Aaij:2014ywa} The results of the measurement, which is performed differentially for three bins in the invariant mass $m_{b \bar b}$ of the $b \bar b$ system, is shown on the right in Figure~\ref{fig:4}. Like in the case of $t \bar t$ production the data shows good agreement with the state-of-the-art SM prediction\,\cite{Gauld:2015qha}  within uncertainties. Another notable feature,\,\cite{Grinstein:2013iws} which is clearly visible in the second bin of the figure, is that the forward-central $b \bar b$ asymmetry $A_{\rm FC}^{b \bar b}$ receives a large correction from purely electroweak effects close to the $Z$ peak. This is not the case for asymmetric $t \bar t$ production, which is fully dominated by QCD effects, making asymmetric $b \bar b$ production an excellent probe of BSM physics entering the electroweak sector.\,\cite{Murphy:2015cha}

\section{Summary and outlook}

LHCb has performed beautiful measurements of a multitude of quark-flavour observables, overall exceeding expectations about its performance and capabilities. Examples include the precision determination of the phase $\phi_s$ in $B_s$--$\bar B_s$ mixing,\,\cite{Aaij:2014zsa} $B_{s (d)} \to \mu^+ \mu^-$, $B \to K^{(\ast)} \mu^+ \mu^-$, $B_s \to \phi \mu^+ \mu^-$, $R_K$, $B \to D^\ast \tau \nu$,\,\cite{Aaij:2015yra} $V_{ub}$  from $\Lambda_b \to p \mu \nu$.\,\cite{Aaij:2015bfa} These results herald the precision era for quark-flavour physics. In some cases the LHCb results pose a serious challenge for theory and improvements are needed to fully exploit existing (future) data. This statement applies in particular to the anomalies seen in the channels $B \to K^{(\ast)} \mu^+ \mu^-$ and  $B_s \to \phi \mu^+ \mu^-$, before one can claim that they are necessarily due to BSM physics. 

There is also a growing and highly interesting LHCb programme beyond standard quark-flavour applications, which unfortunately often does not make it to the front page.  The observation of forward top-quark production, the measurement of the $b \bar b$ forward-central asymmetry or $W$-boson production in association with beauty and charm\,\cite{Aaij:2015cha} are just a few recent examples that resulted from the electroweak physics programme. One can expect more to come in Run~II  from these activities: measurements of the $c \bar c$ charge asymmetry, a precision determination of the $W$-boson mass,\,\cite{Bozzi:2015zja}  maybe even bounds on Higgs production associated with $W/Z$ bosons. 

\section*{Acknowledgements}
I wish to thank the organisers for the invitation to give this plenary talk and also for financial support to attend the conference. I am grateful to Stefan~Dittmaier, Karl~Jakobs and Gavin~Salam for enjoyful discussions.


\begin{thebibliography}{99}

\bibitem{Aad:2012tfa} 
  G.~Aad {\it et al.} [ATLAS Collaboration],
  Phys.\ Lett.\ B {\bf 716}, 1 (2012)
  [arXiv:1207.7214 [hep-ex]].
  
\bibitem{Chatrchyan:2012xdj} 
  S.~Chatrchyan {\it et al.} [CMS Collaboration],
  Phys.\ Lett.\ B {\bf 716}, 30 (2012)
  [arXiv:1207.7235 [hep-ex]].

\bibitem{Aad:2015gba} 
  G.~Aad {\it et al.} [ATLAS Collaboration],
  arXiv:1507.04548 [hep-ex].

\bibitem{CMS:2014ega} 
  CMS Collaboration, 
  CMS-PAS-HIG-14-009.
  
\bibitem{CMS:2014xfa} 
  V.~Khachatryan {\it et al.} [CMS and LHCb Collaborations],
  Nature {\bf 522}, 68 (2015)
  [arXiv:1411.4413 [hep-ex]].
  
\bibitem{Bobeth:2015zqa} 
  C.~Bobeth and U.~Haisch,
  JHEP {\bf 1509}, 018 (2015)
  [arXiv:1503.04829 [hep-ph]].
  
\bibitem{D'Ambrosio:2002ex} 
  G.~D'Ambrosio, G.~F.~Giudice, G.~Isidori and A.~Strumia,
  Nucl.\ Phys.\ B {\bf 645}, 155 (2002)
  [hep-ph/0207036].
  
   \bibitem{Haisch:2007ia} 
  U.~Haisch and A.~Weiler,
  Phys.\ Rev.\ D {\bf 76}, 074027 (2007)
  [arXiv:0706.2054 [hep-ph]].
  
  \bibitem{Ciuchini:2013pca} 
  M.~Ciuchini, E.~Franco, S.~Mishima and L.~Silvestrini,
  JHEP {\bf 1308}, 106 (2013)
  [arXiv:1306.4644 [hep-ph]].
  
 \bibitem{Guadagnoli:2013mru} 
  D.~Guadagnoli and G.~Isidori,
  Phys.\ Lett.\ B {\bf 724}, 63 (2013)
  [arXiv:1302.3909 [hep-ph]].
  
  \bibitem{Bobeth:2005ck} 
  C.~Bobeth, M.~Bona, A.~J.~Buras, T.~Ewerth, M.~Pierini, L.~Silvestrini and A.~Weiler,
  Nucl.\ Phys.\ B {\bf 726}, 252 (2005)
  [hep-ph/0505110].
  
  \bibitem{Hagiwara:1986vm} 
  K.~Hagiwara, R.~D.~Peccei, D.~Zeppenfeld and K.~Hikasa,
  Nucl.\ Phys.\ B {\bf 282}, 253 (1987).

  \bibitem{Hagiwara:1993ck} 
  K.~Hagiwara, S.~Ishihara, R.~Szalapski and D.~Zeppenfeld,
  Phys.\ Rev.\ D {\bf 48}, 2182 (1993).
  
  \bibitem{Altmannshofer:2014rta} 
  W.~Altmannshofer and D.~M.~Straub,
  Eur.\ Phys.\ J.\ C {\bf 75}, no. 8, 382 (2015)
  [arXiv:1411.3161 [hep-ph]].
 
  
  \bibitem{Falkowski:2015jaa} 
  A.~Falkowski, M.~Gonzalez-Alonso, A.~Greljo and D.~Marzocca,
  arXiv:1508.00581 [hep-ph].
  
\bibitem{Brod:2014hsa} 
  J.~Brod, A.~Greljo, E.~Stamou and P.~Uttayarat,
  JHEP {\bf 1502}, 141 (2015)
  [arXiv:1408.0792 [hep-ph]].
    
\bibitem{Rontsch:2014cca} 
  R.~R\"ontsch and M.~Schulze,
  JHEP {\bf 1407}, 091 (2014)
  [arXiv:1404.1005 [hep-ph]].
  
 \bibitem{Descotes-Genon:2013wba} 
  S.~Descotes-Genon, J.~Matias and J.~Virto,
  Phys.\ Rev.\ D {\bf 88}, 074002 (2013)
  [arXiv:1307.5683 [hep-ph]].
  
 \bibitem{Aaij:2013qta} 
  R.~Aaij {\it et al.} [LHCb Collaboration],
  Phys.\ Rev.\ Lett.\  {\bf 111}, 191801 (2013)
  [arXiv:1308.1707 [hep-ex]].
 
 \bibitem{DescotesGenon:2012zf} 
  S.~Descotes-Genon, J.~Matias, M.~Ramon and J.~Virto,
  JHEP {\bf 1301}, 048 (2013)
  [arXiv:1207.2753 [hep-ph]].
  
  \bibitem{LHCb:2015dla} 
  LHCb Collaboration,
  LHCb-CONF-2015-002.
  
  \bibitem{Aaij:2015esa} 
  R.~Aaij {\it et al.} [LHCb Collaboration],
  JHEP {\bf 1509}, 179 (2015)
  [arXiv:1506.08777 [hep-ex]].


  
 \bibitem{Aaij:2014pli} 
  R.~Aaij {\it et al.} [LHCb Collaboration],
  JHEP {\bf 1406}, 133 (2014)
  [arXiv:1403.8044 [hep-ex]].

\bibitem{Horgan:2013pva} 
  R.~R.~Horgan, Z.~Liu, S.~Meinel and M.~Wingate,
  Phys.\ Rev.\ Lett.\  {\bf 112}, 212003 (2014)
  [arXiv:1310.3887 [hep-ph]].
 
\bibitem{Straub:2015ica} 
  A.~Bharucha, D.~M.~Straub and R.~Zwicky,
  arXiv:1503.05534 [hep-ph].
  
  \bibitem{Aaij:2014ora} 
  R.~Aaij {\it et al.} [LHCb Collaboration],
  Phys.\ Rev.\ Lett.\  {\bf 113}, 151601 (2014)
  [arXiv:1406.6482 [hep-ex]].
  
  \bibitem{LHCb14talk}
  U.~Haisch, talk at ``Implications of LHCb measurements \& future prospects", 17 October~2014, \href{https://indico.cern.ch/event/324660/session/7/contribution/50/attachments/629120/865763/LHCb14.pdf}{https://indico.cern.ch/event/324660/session/7/contribution/50/attachments/ 629120/865763/LHCb14.pdf}
  
  
  \bibitem{Altmannshofer:2015sma} 
  W.~Altmannshofer and D.~M.~Straub,
  arXiv:1503.06199 [hep-ph].
  
  \bibitem{Descotes-Genon:2015uva} 
  S.~Descotes-Genon, L.~Hofer, J.~Matias and J.~Virto,
  arXiv:1510.04239 [hep-ph].
  
  \bibitem{Khodjamirian:2010vf} 
  A.~Khodjamirian, T.~Mannel, A.~A.~Pivovarov and Y.-M.~Wang,
  JHEP {\bf 1009}, 089 (2010)
  [arXiv:1006.4945 [hep-ph]].
  
 \bibitem{Descotes-Genon:2014uoa} 
  S.~Descotes-Genon, L.~Hofer, J.~Matias and J.~Virto,
  JHEP {\bf 1412}, 125 (2014)
  [arXiv:1407.8526 [hep-ph]].
  
  \bibitem{Lyon:2014hpa} 
  J.~Lyon and R.~Zwicky,
  arXiv:1406.0566 [hep-ph].
  
  \bibitem{Gauld:2013qja} 
  R.~Gauld, F.~Goertz and U.~Haisch,
  JHEP {\bf 1401}, 069 (2014)
  [arXiv:1310.1082 [hep-ph]].
  
  \bibitem{Buras:2013dea} 
  A.~J.~Buras, F.~De Fazio and J.~Girrbach,
  JHEP {\bf 1402}, 112 (2014)
  [arXiv:1311.6729 [hep-ph]].
  
  \bibitem{inpreparation} 
  R.~Gauld, F.~Goertz and U.~Haisch,
  in preparation.
  
  \bibitem{Altmannshofer:2014cfa} 
  W.~Altmannshofer, S.~Gori, M.~Pospelov and I.~Yavin,
  Phys.\ Rev.\ D {\bf 89}, 095033 (2014)
  [arXiv:1403.1269 [hep-ph]].
  
  \bibitem{Crivellin:2015mga} 
  A.~Crivellin, G.~D'Ambrosio and J.~Heeck,
  Phys.\ Rev.\ Lett.\  {\bf 114}, 151801 (2015)
  [arXiv:1501.00993 [hep-ph]].
  
  \bibitem{Crivellin:2015lwa} 
  A.~Crivellin, G.~D'Ambrosio and J.~Heeck,
  Phys.\ Rev.\ D {\bf 91}, no. 7, 075006 (2015)
  [arXiv:1503.03477 [hep-ph]].
  
  \bibitem{Aad:2014cka} 
  G.~Aad {\it et al.} [ATLAS Collaboration],
  Phys.\ Rev.\ D {\bf 90}, no. 5, 052005 (2014)
  [arXiv:1405.4123 [hep-ex]].
  
  \bibitem{Coco}
  V.~Coco on behalf of the LHCb Collaboration, LHC Seminar, 26 May 2015, \href{https://indico.cern.ch/event/388144/attachments/775827/1063900/Top_Coco.pdf}{https://indico.cern.ch/event/388144/attachments/ 775827/1063900/Top\_Coco.pdf}
  
  \bibitem{Aaij:2015mwa} 
  R.~Aaij {\it et al.} [LHCb Collaboration],
  Phys.\ Rev.\ Lett.\  {\bf 115}, no. 11, 112001 (2015)
  [arXiv:1506.00903 [hep-ex]].
  
  \bibitem{Kagan:2011yx} 
  A.~L.~Kagan, J.~F.~Kamenik, G.~Perez and S.~Stone,
  Phys.\ Rev.\ Lett.\  {\bf 107}, 082003 (2011)
  [arXiv:1103.3747 [hep-ph]].
  
  \bibitem{Gauld:2014pxa} 
  R.~Gauld,
  Phys.\ Rev.\ D {\bf 91}, 054029 (2015)
  [arXiv:1409.8631 [hep-ph]].
  
  \bibitem{Gauld:2013aja} 
  R.~Gauld,
  JHEP {\bf 1402}, 126 (2014)
  [arXiv:1311.1810 [hep-ph]].
  
  \bibitem{Aaij:2014ywa} 
  R.~Aaij {\it et al.} [LHCb Collaboration],
  Phys.\ Rev.\ Lett.\  {\bf 113}, no. 8, 082003 (2014)
  [arXiv:1406.4789 [hep-ex]].
  
  \bibitem{Gauld:2015qha} 
  R.~Gauld, U.~Haisch, B.~D.~Pecjak and E.~Re,
  Phys.\ Rev.\ D {\bf 92}, 034007 (2015)
  [arXiv:1505.02429 [hep-ph]].
  
  \bibitem{Grinstein:2013iws} 
  B.~Grinstein and C.~W.~Murphy,
  Phys.\ Rev.\ Lett.\  {\bf 111}, 062003 (2013)
  [Phys.\ Rev.\ Lett.\  {\bf 112}, no. 23, 239901 (2014)]
  [arXiv:1302.6995 [hep-ph]].
  
  \bibitem{Murphy:2015cha} 
  C.~W.~Murphy,
  Phys.\ Rev.\ D {\bf 92}, no. 5, 054003 (2015)
  [arXiv:1504.02493 [hep-ph]].
  
  \bibitem{Aaij:2014zsa} 
  R.~Aaij {\it et al.} [LHCb Collaboration],
  Phys.\ Rev.\ Lett.\  {\bf 114}, no. 4, 041801 (2015)
  [arXiv:1411.3104 [hep-ex]].
  
  \bibitem{Aaij:2015yra} 
  R.~Aaij {\it et al.} [LHCb Collaboration],
  Phys.\ Rev.\ Lett.\  {\bf 115}, no. 11, 111803 (2015)
  [arXiv:1506.08614 [hep-ex]].
  
  \bibitem{Aaij:2015bfa} 
  R.~Aaij {\it et al.} [LHCb Collaboration],
  Nature Phys.\  {\bf 11}, 743 (2015)
  [arXiv:1504.01568 [hep-ex]].
  
  \bibitem{Aaij:2015cha} 
  R.~Aaij {\it et al.} [LHCb Collaboration],
  Phys.\ Rev.\ D {\bf 92}, no. 5, 052001 (2015)
  [arXiv:1505.04051 [hep-ex]].
  
  \bibitem{Bozzi:2015zja} 
  G.~Bozzi, L.~Citelli, M.~Vesterinen and A.~Vicini,
  arXiv:1508.06954 [hep-ex].
  
  

\end{thebibliography}
\end{document}